\documentclass[11pt]{article}
\usepackage{dcolumn}
\usepackage{bm}
\usepackage{amsmath}
\usepackage{amssymb}
\usepackage{amscd}
\usepackage{afterpage}
\usepackage{float,times}
\usepackage{subfigure}
\usepackage{rotating}
\usepackage{multirow}
\usepackage{epsfig}
\usepackage{theorem}
\usepackage{moreverb}
\usepackage{euscript}
\usepackage{psfrag}

\begin{document}

\begin{titlepage}
\begin{center}
{\hbox to\hsize{\hfill June 2009 }}

\bigskip

\bigskip

\bigskip

\vspace{6\baselineskip}

{\Large \bf
On the infrared limit of Ho{\v r}ava's gravity \\ with the global Hamiltonian constraint\\
}
\bigskip

\bigskip

{\bf Archil Kobakhidze  \\}
\smallskip

{ \small \it 
School of Physics, The University of Melbourne, Victoria 3010, Australia \\ 
E-mail: archilk@unimelb.edu.au \\}

\bigskip

\vspace*{.5cm}

{\bf Abstract}\\
\end{center}

\noindent
{\small 
We show that Ho{\v r}ava's theory of gravitation with the global Hamiltonian  constraint does not reproduce General Relativity in the infrared domain. There is one extra propagating degree of freedom, besides those two associated with the massless graviton, which does not decouple. }  

\bigskip

\bigskip

\bigskip

\end{titlepage}

\paragraph{Introduction.}

Recently, Ho{\v r}ava proposed a power-counting renormalizable higher-derivative theory of  gravitation where the full diffeomorphism invariance is broken down to the foliation-preserving diffeomorphism \cite{Horava:2009uw}. Because of the reduced symmetry, the ghost states usually associated with the higher time derivatives in General Relativity (GR), are removed, and thus the theory is unitary. A vital question is whether Ho{\v r}ava's theory of gravitation has an infrared limit consistent with observations. Since the observational success of GR is largely based on its full diffeomorphism invariance, it is clear that any theory with reduced diffeomorphism invariance will deviate from GR in both the ultraviolet and infrared regimes. From the purely phenomenological point of view, it is important to understand whether this deviation in the infrared regime can be made consistent with observations. In \cite{Charmousis:2009tc} it has been pointed out that the scalar polarization of the graviton does not decouple in Ho{\v r}ava's theory. Also, it has been shown in \cite{Li:2009bg} that in Ho{\v r}ava's theory with local Hamiltonian constraint the Poisson algebra is not closed. From this perspective it seems vital to retain the "projectability condition" which generates a less restrictive global Hamiltonian constraint \cite{Mukohyama:2009mz}. In this paper we consider the infrared limit of Ho{\v r}ava's theory with the global Hamiltonian constraint. By applying Dirac's constraint analysis, we show that there is one extra propagating degree of freedom, besides those two associated with the massless graviton. Therefore, Ho{\v r}ava's theory of gravitation does not reproduce General Realativity in the infrared regime.

\paragraph{Dirac's constraint analysis.}

In the infrared limit, Ho{\v r}ava's theory is described by the following action\footnote{We adopt units where $16\pi G_N=1$}:
\begin{equation}
S_{\rm Horava}=\int dt\int_{\Sigma_{t}}d^3x\left(\pi^{ij}\dot h_{ij}-N\mathcal{H}_{0}-N_i\mathcal{H}^{i}\right)~, 
\label{1}
\end{equation}
where
\begin{equation}
\mathcal{H}_{0}=-\sqrt{h}~ ^{(3)}R+h^{-1/2}\pi^{ij}\pi_{ij}-\frac{\lambda}{(3\lambda -1)}h^{-1/2}\pi^2~,
\label{2}
\end{equation} 
\begin{equation}
\mathcal{H}^i=-2\nabla_i(\pi^{ij})~,
\label{3}
\end{equation} 
and $\pi^{ij}$ is a canonical momentum for the spatial metric $h_{ij}$ defined on a spatial hypersurface $\Sigma_t$ [$h\equiv {\rm det}(h_{ij})$]. $^{(3)}R$ denotes the Ricci scalar build up from the metric $h_{ij}$, and $\nabla_i$ is the covariant derivative associated with $h_{ij}$. It has been argued (but has not been explicitly shown) in \cite{Horava:2009uw}  that the effective running parameter $\lambda$ has an infrared fixed point at $\lambda=1$. In what follows we also assume that $\lambda=1$. We also set to 0 the cosmological constant in the original theory \cite{Horava:2009uw}. This seems in principle possible if the "detailed balance condition" of the original theory is abandoned \cite{Sotiriou:2009gy}. Anyway, the cosmological constant is not important for our analyses. With these simplifying assumptions, the action (\ref{1}) looks identical to the GR action with one important exception: the lapse function depends only on time variable, $N=N(t)$. Therefore, the action (\ref{1}) is invariant under the reduced diffeomorphism symmetry, the foliation-preserving diffeomorphism transformations.  

Since Ho{\v r}ava's theory maintains a smaller diffeomorphism group, an important question is, how many degrees of freedom does the theory describe? This question can be answered by applying Dirac's constraint analysis. We define canonical momenta: $\pi_N\equiv\frac{\partial \mathcal{L}_{\rm Horava}}{\partial \dot N}\approx0$, $\pi_N^i\equiv\frac{\partial \mathcal{L}_{\rm Horava}}{\partial \dot N_i}\approx 0$ and $\pi^{ij}\equiv\frac{\partial \mathcal{L}_{\rm Horava}}{\partial \dot h_{ij}}$. The basic Poisson commutators read:
\begin{equation}
\left\lbrace N, \pi_N \right\rbrace_{\rm PB}=1~, 
\label{3a}
\end{equation}
\begin{equation}
\left\lbrace N_i(x), \pi_N^j(y) \right\rbrace_{\rm PB}=\delta_i^j\delta^{3}(x-y)~, 
\label{3b}
\end{equation}
and 
\begin{equation}
\left\lbrace h_{ij}(x), \pi^{mn}(y) \right\rbrace_{\rm PB}=\frac{1}{2}\left(\delta_i^m\delta_j^n+\delta_i^n\delta_j^m\right)\delta^{3}(x-y)~. 
\label{3c}
\end{equation}
Note that, because the lapse function depends only on the time coordinate, its canonical momentum $\pi_N$ is also $x$-independent. Consequently, the corresponding secondary Hamiltonian constraint is also satisfied only for the zero mode of $\mathcal{H}_{0}$. Indeed, taking the Hamiltonian
\begin{equation}
H=N\int_{\Sigma_t} d^3x \mathcal{H}_0+\int_{\Sigma_t} d^3x N^i\mathcal{H}_i~,
\label{3d}
\end{equation}
one easily finds
\begin{equation}
\dot\pi_N=\left\lbrace \pi_N, H \right\rbrace_{\rm PB}=\int_{\Sigma_t} d^3y \mathcal{H}_0(y)~.
\label{3e}
\end{equation}
That is, to preserve the primary constraint $\pi_N\approx 0$ in time, one must assume, 
\begin{equation}
H_0\equiv \int_{\Sigma_t}d^3x \mathcal{H}_{0}\approx 0~.
\label{4}
\end{equation}
The secondary momentum constraints, on the other hand, are satisfied locally, i.e. at each given point $x$: 
\begin{equation}
\mathcal{H}^{i}(x)\approx 0~,
\label{5}
\end{equation}
just like in GR \cite{DeWitt:1967yk}. Next, we compute the algebra of secondary constraints:
\begin{eqnarray}
\left\lbrace H_0,H_0 \right\rbrace_{\rm PB}=\int\int d^3xd^3y\left(2\mathcal{H}^{i}(x)\partial^{(x)}_{i}\delta^3(x-y)
+\partial_{i}^{(x)}\mathcal{H}^i(x)\delta^3(x-y)\right) \nonumber \\
=\int d^3x \partial_{i}^{(x)}\mathcal{H}^i(x)=0~, 
\label{6}
\end{eqnarray}
where we have used $\int d^3x \partial^{(x)}_{i}\delta^3(x)=0$ and $\mathcal{H}^{i}\stackrel{|x|\to \infty}{\longrightarrow}0$;
\begin{eqnarray} 
\left\lbrace {\cal H}_i(x), H_0 \right\rbrace_{\rm PB}=\mathcal{H}_0(x)\partial^{(x)}_i\int d^3y \delta^{3}(x-y)=0~; 
\label{7}
\end{eqnarray}
and the last commutator,
\begin{eqnarray}
\left\lbrace {\cal H}_i(x),{\cal H}_j(y) \right\rbrace_{\rm PB}={\cal H}_i(y)\partial^{(y)}_j\delta^3(x-y)-{\cal H}_j(x) \partial^{(x)}_i\delta^{3}(y-x)\approx 0~.
\label{8}
\end{eqnarray}
is the same as the one in GR \cite{DeWitt:1967yk}. Using the above relations we verify that no further constraints emerge:
\begin{equation}
\dot H_0=\left\lbrace H_0, H\right\rbrace_{\rm PB}=0~,~~\dot {\cal H}_i=\left\lbrace {\cal H}_i, H\right\rbrace_{\rm PB}\approx 0~.
\label{9}
\end{equation}

The set of constraints $\left\lbrace \pi_N, \pi_N^i, H_0, {\cal H}^i\right\rbrace$ for zero modes, and the set of constraints $\left\lbrace \pi_N^i, {\cal H}^i\right\rbrace$ for propagating modes represent non-singular systems of the first-class constraints. Therefore, we can directly apply Dirac's method of counting the physical degrees of freedom:
\begin{eqnarray}
\left[\text{number of physical d.o.f }\right]=\frac{1}{2}\left[\text{number of canonical variables}\right]-
\left[\text{number of first-class constraints}\right] 
\label{9a}
\end{eqnarray} 
Applying the above counting to non-propagating ($x$-independent) zero modes we obtain: $10-8=2$, similar to GR. For propagating modes, however, we have one extra physical degree of freedom: $9-6=3$. The difference in global and local degrees of freedom seems to be related with the non-local nature of the Ho{\v r}ava gravity where the lapse function strictly depends only on the time coordinate.

It is instructive to compare the counting of degrees of freedom in Ho{\v r}ava's theory with the counting in other 
theories with reduced diffeomorphism invariance. Namely, in the unimodular theory of gravitation only coordinate transformations with unit Jacobian are admissible. In that case, $(h^{1/2}N)$ is fixed and, hence, there is no Hamiltonian constraint. Instead, there are  first-class tertiary constraints, $\partial_i(h^{-1/2}{\cal H}_0)=0$ \cite{Henneaux:1989zc}. These tertiary constraints can be solved by defining a new Hamiltonian constrain, ${\cal H}^{\prime}_{0}\equiv{\cal H}_0+h^{1/2}\Lambda\approx 0$, where $\Lambda$ is a zero-mode field, $\partial_i\Lambda=0$. The conservation of this Hamiltonian constraint then implies that $\Lambda$ is actually a 'vacuum' field, $\Lambda={\rm const}$. Thus, besides the two degrees of freedom associated with massless graviton, one accounts in addition literally one (not per each point $x$) global degree of freedom. This global degree of freedom turns out to be a cosmological constant and the "cosmic" time is its canonically conjugated variable. Consequently, one can rewrite the unimodular theory as an equivalent fully covariant theory by reparameterizing the time coordinate \cite{Henneaux:1989zc} without introducing new propagating degrees of freedom. A similar trick seems  impossible in Ho{\v r}ava's theory, because the extra degree of freedom is a propagating mode. That is to say, to covariantize Ho{\v r}ava's gravity one necessarily needs to invoke a new dynamical field.\footnote{Such a covariant theory has been proposed recently in \cite{Germani:2009yt}. However, it is assumed there that the lapse function is a fully-fledged field. It has been claimed also that for $\lambda=1$ the extra degree of freedom "freezes out". We note here that the set of constraints introduced in \cite{Germani:2009yt} seems to be singular, and thus the counting of degrees of freedom must be taken with care. }

\paragraph{A covariant action.}

In this section we would like to write down an equivalent to (\ref{1}) action where the lapse function is promoted to a full space-time dependent field, $N= N(x,t)$. The action (\ref{1}) turns then into a Einstein-Hilbert action. The  projectability condition on the lapse function can be enforced through the equation of motion, $\partial_i N=0$. In order to achieve this in a covariant way, we introduce a spatial two-form field ${\cal A}_{ij}$, whose field strength is ${\cal F}_{ijk}=\partial_{\left[i\right.}{\cal A}_{\left. jk\right]}$. The dual field strength then represents spatial density, 
\begin{equation}
\tilde {\cal F}\equiv h^{1/2}\frac{1}{3!}\epsilon^{ijk}{\cal F}_{ijk}=\partial_i\tilde {\cal A}^i~,
\label{10}
\end{equation}
where $\tilde {\cal A}^i=h^{1/2}\epsilon^{ijk}{\cal A}_{jk}$ is a 3-vector density. The diffeomorphism invariant action equivalent to (\ref{1}) then takes the form:
\begin{equation}
S_{\rm equiv.}=S_{\rm GR}+\int d^4x N\partial_i\tilde {\cal A}^i~.
\label{11}
\end{equation} 
 where $S_{\rm GR}$ is the standard Einstein-Hilbert action. Varying the above action with respect to $\tilde {\cal A}^i$ we obtain the desired constraint equation
\begin{equation}
\partial_i N=0~.
\label{12}
\end{equation} 
Also, the action (\ref{12}) implies the local Hamiltonian constraint, 
\begin{equation}
{\cal H}_0+\partial_i\tilde {\cal A}^i\approx 0~.
\label{13}
\end{equation}
Integrating the above equation over the constant time hypersurface, and assuming $\tilde {\cal A}^i\stackrel{|x|\to \infty}{\longrightarrow}0$ we reproduce the global Hamiltonian constraint (\ref{4}). 

Observe now  that the modified Hamiltonian constraint (\ref{13}) together with the momentum constraint equations and the dynamical equations of motion form the following set of Einstein's equations:
\begin{equation}
^{(4)}G_{\mu\nu}=-Fn_{\mu}n_{\nu}~,
\label{13a}
\end{equation}
where $n_{\mu}=(1,0,0,0,)$ and $^{(4)}G_{\mu\nu}$ is the Einstein tensor built from 4D metric tensor $^{(4)}g_{\mu\nu}$. 
The additional term in (\ref{11}) can be viewed as the  action functional of a pressureless dust, where $F=2\tilde {\cal F}/h^{1/2}$ is the energy density of dust particles as seen by observers who are at rest in the constant time hypersurfaces. This observation has been first made in \cite{Mukohyama:2009mz}. However, this is not the ordinary dust fluid as its total energy is zero, 
\begin{equation}
\int d^3x \tilde {\cal F} = 0~.
\label{14}
\end{equation}  
Note that the "free" (vacuum) limit of equations (\ref{13a}) can not be achieved because, according to eq. (\ref{13}), $F(x)=0$ would correspond to the local Hamiltonian constraint for ${\cal H}_0$ which is not admissible in Ho{\v r}ava's gravity. In fact, taking the trace of (\ref{13a}) one can solve for the non-dynamical auxiliary field $F$, 
\begin{equation}
F=\frac{1}{g^{00}}R~.
\label{13b}
\end{equation}
Substituting (\ref{13b}) back into (\ref{13a}) we obtain the following equations 
\begin{equation}
R_{\mu\nu}-\left(\frac{1}{2}g_{\mu\nu}-\frac{1}{g^{00}}n_{\mu}n_{\nu}\right)R = 0~.
\label{13c}
\end{equation}
This equations can be viewed as the vaccum equations of the theory. They must be suplemented by the condition (\ref{12}), where  $N=(-g^{00})^{-1/2}$. Applying the contracted Bianchi identities to (\ref{13c}) one finds, $\partial^0 R = 0$, which in turn implies $R=\rho(\vec x)$, where $\rho(\vec x)$ is a function of spatial coordinates only. Therefore, the spatial dependence of $F$ (\ref{13c}) is entirely determined by this function, $F=\frac{1}{g^{00}}\rho$, and the global Hamiltonian constraint reads as: $\int d^3x h^{1/2}\rho = 0$. Finally, the coupling to a matter energy-momentum tensor $T_{\mu\nu}$ is described by adding $\frac{1}{2}\left(T_{\mu\nu}-T^{\alpha}_{\alpha}\frac{n_{\mu}n_{\nu}}{g^{00}}\right)$ to the rhs of eq. (\ref{13c}). 

\paragraph{The weak-field limit}

Let us now consider the linearized version of (\ref{13c}), by expanding the metric around flat Minkowski background , $\eta_{\mu\nu}$, 
\begin{equation}
^{(4)}g_{\mu\nu}\approx \eta_{\mu\nu}+\gamma_{\mu\nu}~.
\label{15}
\end{equation}
This expansion is justified, providing one considers $F$ in (\ref{13b}) as a small perturbation, $F\approx -\rho\sim {\cal O}(h)$.   

We choose to work in the temporal gauge, $\gamma_{\mu 0}=0$. In this gauge the constraint equation (\ref{12}) is automatically satisfied. The above gauge fixing conditions are preserved by the residual diffeomorphism transformations with $\xi_0\equiv \xi(\vec x),~\xi_i\equiv \xi_{i}(\vec x)+t\xi_0(\vec x)$. In \cite{Horava:2009uw} this residual invariance has been used to impose further conditions, $\partial^{i}\gamma_{ij}-\partial_j\gamma^{i}_{i}=0$. However, these conditions imply immediately, ${\cal H}_0\approx \partial^{j}\left(\partial^{i}\gamma_{ij}-\partial_j\gamma^{i}_{i} \right)+{\cal O}(\gamma^2)=0$, and therefore they are not admissible in the theory without a local Hamiltonian constraint. That is to say, the absence of the local Hamiltonian constraint prevents us from removing an extra (compared to GR) propagating degree of freedom, in full accordance with our generic Dirac constraint analysis.  

To see how this extra degree of freedom affects the gravitational interactions between matter sources, let us more closely inspect the linearized equations amended by the conserved matter energy-momentum tensor $T_{\mu\nu}$ ($\partial^{\mu}T_{\mu\nu}=0$): 
\begin{eqnarray}
\Box \gamma_{\mu\nu}-\partial_{\mu}\partial^{\alpha}\gamma_{\alpha \nu}-
\partial_{\nu}\partial^{\alpha}\gamma_{\mu\alpha}+\partial_{\mu\nu}\gamma 
+(\eta_{\mu\nu}+2n_{\mu}n_{\nu})\left(\partial^{\alpha}\partial^{\beta}-\Box \gamma\right)=-2\left(T_{\mu\nu}+n_{\mu}n_{\nu}T\right)
\label{15a}
\end{eqnarray}
where $\Box \equiv \partial^{\alpha}\partial_{\alpha}$, $\gamma\equiv \gamma^{\alpha}_{\alpha}$ and $T\equiv T^{\alpha}_{\alpha}$. The solution to (\ref{15a}) can be written as:
\begin{equation}
\gamma_{\mu\nu}=\tilde \gamma_{\mu\nu}+\gamma_{\mu\nu}^{\rm GR}~,
\label{15b}
\end{equation} 
where $\gamma_{\mu\nu}^{\rm GR}$ is the solution of the corresponding Einstein's equation:
\begin{equation}
\gamma_{\mu\nu}^{\rm GR}=\int dy G_{\mu\nu\rho\sigma}^{\rm GR}(x-y)T^{\rho\sigma}(y)~.
\label{15c}
\end{equation}
$G_{\mu\nu\rho\sigma}^{\rm GR}$ in the above equation is the GR causal graviton propagator in the temporal gauge, $n^{\mu}G_{\mu\nu\rho\sigma}^{\rm GR}=n^{\rho}G_{\mu\nu\rho\sigma}^{\rm GR}=0$. $\tilde \gamma_{\mu\nu}$ in (\ref{15b}) satisfies the homogeneous equation, 
\begin{equation}
\Box \tilde \gamma_{\mu\nu}-\partial_{\mu}\partial^{\alpha}\tilde \gamma_{\alpha \nu}-
\partial_{\nu}\partial^{\alpha}\tilde \gamma_{\mu\alpha}+\partial_{\mu\nu}\tilde \gamma 
+(\eta_{\mu\nu}+2n_{\mu}n_{\nu})\left(\partial^{\alpha}\partial^{\beta}-\Box \tilde \gamma\right)=0~,
\label{15d}
\end{equation}
and, thus, contributes to the on-shell part of the total propagator, modifying normal analytic properties which characterize the standard causal propagators. 
Indeed, the standard causal  propagator is determined (up to a tensorial part) by the pole structure in the momentum space, $G(p) \varpropto (p^2-i\epsilon)^{-1}={\rm P.V.} p^{-2}-i\pi\delta(p^2)$. This propagator is obtained by prescribing appropriate asymptotic boundary conditions at $t=\pm infty$. Assuming that, $G_{\mu\nu\rho\sigma}^{\rm GR}$ in (\ref{15c}) is such a causal propagator, it is easy to see that the total propagator in the Ho{\v r}ava gravity will contain additional, Lorentz non-invariant, on-shell piece due to the contribution from $\tilde \gamma_{\mu\nu}$ obeying (\ref{15d}):   $G_{\mu\nu\rho\sigma}^{\rm GR}(p)\varpropto {\rm P.V.} p^{-2}-i\pi(1+f(p))\delta(p^2) $, where $f(p_0, p_i=0)=0$.  Since the pole structure of the propagator is ultimately related to the unitarity and causality, the graviton exchange amplitudes in Ho{\v r}ava's theory are likely to fail to satisfy unitarity/causality conditions. This is the consequence of the modified asymptotic boundary conditions in the Ho{\v r}ava gravity due to the non-decoupling of the extra scalar mode which has been established rigorously, without recourse to a background metric and the linearized approximation, in the previous sections.

The discontinuity of an apparent GR limit is typical in theories with broken diffeomorphism invariance. The well-known example is the Pauli-Fierz theory of massive gravity. The massless limit of massive gravity is known to be discontinuous \cite{vanDam:1970vg}, \cite{Zakharov:1970cc}, and does not coincide with GR, at least at perturbative level. The key reason is, of course, breaking of the diffeomorphism invariance by the graviton mass, so massless and massive theories describe a different number of degrees of freedom. In particular, the scalar graviton does not decoupled in the massless limit. Similarly, in Ho{\v r}ava's theory with a global Hamiltonian constraint, breaking of the full diffeomorphism invariance results in propagating a massless scalar graviton. Moreover, contrary to the massive gravity \cite{Kogan:2000uy}, \cite{Porrati:2000cp}, the problem of extra unwanted degree of freedom seems to persist in Ho{\v r}ava's theory with the cosmological constant, because the dust acts not like a mass, but as a source which cannot be switched off. 

In conclusion, Ho{\v r}ava's theory of gravitation with the global Hamiltonian constraint in the infrared regime contains an extra propagating massless degree of freedom which can not be removed. This, together with other negative results reported in the literature, puts serious doubt on the validity of the theory.  

\subparagraph{Acknowledgments.}

I am indebted to Ray Volkas for discussions on various aspects of Ho{\v r}ava's gravity. The work was supported by the Australian Research Council

\subparagraph{Note added.} 
During the preparation of this paper, Ref. \cite{Blas:2009yd} appeared on the hep-th archive, where some important issues related with the extra degree of freedom in Ho{\v r}ava's theory with a local Hamiltonian constraint has been clarified. They have also briefly discussed the version of Ho{\v r}ava's theory with global Hamiltonian constraint, by  associating it with a covariant theory admitting a ghost condensate. This treatment is different from the one discussed in the present paper.
  



\begin{thebibliography}{99}


\bibitem{Horava:2009uw}
  P.~Ho\v{r}ava,
  ``Quantum Gravity at a Lifshitz Point,''
  Phys.\ Rev.\  D {\bf 79} (2009) 084008
  [arXiv:0901.3775 [hep-th]];
  P.~Ho{\v r}ava,
  ``Membranes at Quantum Criticality,''
  JHEP {\bf 0903} (2009) 020
  [arXiv:0812.4287 [hep-th]].

\bibitem{Charmousis:2009tc}
  C.~Charmousis, G.~Niz, A.~Padilla and P.~M.~Saffin,
  ``Strong coupling in Ho{\v r}ava gravity,''
  arXiv:0905.2579 [hep-th].
  
\bibitem{Li:2009bg}
  M.~Li and Y.~Pang,
  ``A Trouble with Ho\v{r}ava-Lifshitz Gravity,''
  arXiv:0905.2751 [hep-th].

\bibitem{Sotiriou:2009gy}
  T.~Sotiriou, M.~Visser and S.~Weinfurtner,
  ``Phenomenologically viable Lorentz-violating quantum gravity,''
  arXiv:0904.4464 [hep-th].
  
\bibitem{Mukohyama:2009mz}
  S.~Mukohyama,
  ``Dark matter as integration constant in Ho{\v r}ava-Lifshitz gravity,''
  arXiv:0905.3563 [hep-th].
  
\bibitem{DeWitt:1967yk}
  B.~S.~DeWitt,
  ``Quantum Theory of Gravity. 1. The Canonical Theory,''
  Phys.\ Rev.\  {\bf 160} (1967) 1113.


\bibitem{Henneaux:1989zc}
  M.~Henneaux and C.~Teitelboim,
  ``The cosmological constant and general covariance,''
  Phys.\ Lett.\  B {\bf 222} (1989) 195.


\bibitem{Germani:2009yt}
  C.~Germani, A.~Kehagias and K.~Sfetsos,
  ``Relativistic Quantum Gravity at a Lifshitz Point,''
  arXiv:0906.1201 [hep-th].

\bibitem{vanDam:1970vg}
  H.~van Dam and M.~J.~G.~Veltman,
  ``Massive And Massless Yang-Mills And Gravitational Fields,''
  Nucl.\ Phys.\  B {\bf 22} (1970) 397.
  
\bibitem{Zakharov:1970cc}
  V.~I.~Zakharov,
  ``Linearized gravitation theory and the graviton mass,''
  JETP Lett.\  {\bf 12} (1970) 312
  [Pisma Zh.\ Eksp.\ Teor.\ Fiz.\  {\bf 12} (1970) 447].



\bibitem{Kogan:2000uy}
  I.~I.~Kogan, S.~Mouslopoulos and A.~Papazoglou,
  ``The m $\to$ 0 limit for massive graviton in dS(4) and AdS(4): How to
   circumvent the van Dam-Veltman-Zakharov discontinuity,''
  Phys.\ Lett.\  B {\bf 503} (2001) 173
  [arXiv:hep-th/0011138].

\bibitem{Porrati:2000cp}
  M.~Porrati,
  ``No van Dam-Veltman-Zakharov discontinuity in AdS space,''
  Phys.\ Lett.\  B {\bf 498} (2001) 92
  [arXiv:hep-th/0011152].

\bibitem{Blas:2009yd}
  D.~Blas, O.~Pujolas and S.~Sibiryakov,
  ``On the Extra Mode and Inconsistency of Ho{\v r}ava Gravity,''
  arXiv:0906.3046 [hep-th].


  
\end{thebibliography}
\end{document}